# THE IMPORTANCE OF HISTORICAL MEASURES FOR DYNAMICAL MODELS OF THE EVOLUTION OF TRAPEZIUM-TYPE MULTIPLE SYSTEMS

**Christine Allen, Leonardo J. Sánchez, Alex Ruelas-Mayorga and Rafael Costero**

*Instituto de Astronomía, Universidad Nacional Autónoma de México, Ciudad de México, 04510, México.*
E-mails: allen@astro.unam.mx, leonardo@astro.unam.mx, rarm@astro.unam.mx, costero@astro.unam.mx

**Abstract:** As an illustration of the value of historical measures we present some examples of the dynamical evolution of multiple systems resembling the Orion Trapezium. We constructed models by combining carefully selected historical measures of the separations among components of young massive stellar systems with modern observations. By computing large numbers of fictitious systems resembling real trapezia we were able to simulate the dynamical evolution of such systems. Our results on the dynamical fate of the Orion Trapezium and of ten additional young clusters resembling the Orion Trapezium show extremely short dynamical lifetimes for these systems.

## 1 INTRODUCTION

Massive stars are currently thought to be formed in small clusters, similar to the Orion Trapezium. According to the definition given by Ambartsumian (1954), a multiple star (with three or more stars) is of trapezium type if three or more distances between components are of the same order of magnitude. "Of the same order of magnitude" means, in this context, that their ratio be greater than ⅓ but less than 3. If no three distances are of the same order of magnitude, then the system is of ordinary type, according to Ambartsumian. For the sake of clarity, we shall call the latter 'hierarchical systems'.

Trapezium-type clusters—or multiple systems—have been shown to be dynamically very unstable, because their configurations (similar separations among their components) rapidly lead to close encounters which result in one or more members being expelled from the system. Gas loss shortly after the stars are formed also contributes to the instability of such multiple systems. The aim of this paper is to highlight the manner in which historical measures can be used to gain understanding about the internal motions of multiple systems. These data can then be used to construct models of their dynamical evolution.

There exists a large number of historical observations of position angles and separations among components of multiple stars, available from the United States Naval Observatory (USNO). Many of these measures were obtained visually by observers who were able to achieve surprisingly accurate results mostly using filar micrometers. A comprehensive historical review of double star observers is given by Tenn (2013).

Carefully selected historical measures of the separations and position angles are valuable because they can be combined with modern measures, thus providing a long time baseline not otherwise available. This enables us to obtain the relative velocities of the components in the plane of the sky. In the past, we have successfully used a combination of historical and modern measures to gain insight into the internal motions of some trapezia (Allen et al., 1974; 2004). More recently, we have expanded these studies to include new data on the Orion Trapezium and on other massive trapezia, and to construct models of their dynamical evolution (Allen et al., 2015; 2017; 2018).

Our present aim is to emphasise that historical observations can be combined with modern ones to yield valuable insights into the dynamical evolution of multiple systems such as the Orion Trapezium. As an illustration of the use of historical measures to obtain such insight into the early dynamical evolution of young stellar groups, we present some results of our models of the Orion Trapezium itself and of ten additional massive multiple systems resembling this Trapezium.

## 2 SELECTION OF HISTORICAL OBSERVATIONS

Since we combine historical and modern observations in order to have a long time baseline, it is important to select only old measures with a precision comparable with that of modern ones. As pointed out in Olivares et al. (2013), separation measures of selected observers are usually more reliable than position angle measures. Therefore, we neglected the measures of position angles and considered only historical separation measures by the 'best' observers, as described in Allen et al. (1974; 2004).

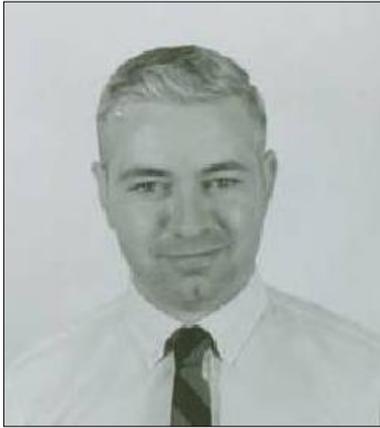

Figure 1: Charles Edmund Worley in 1964, at the USNO (after Mason et al., 2007).

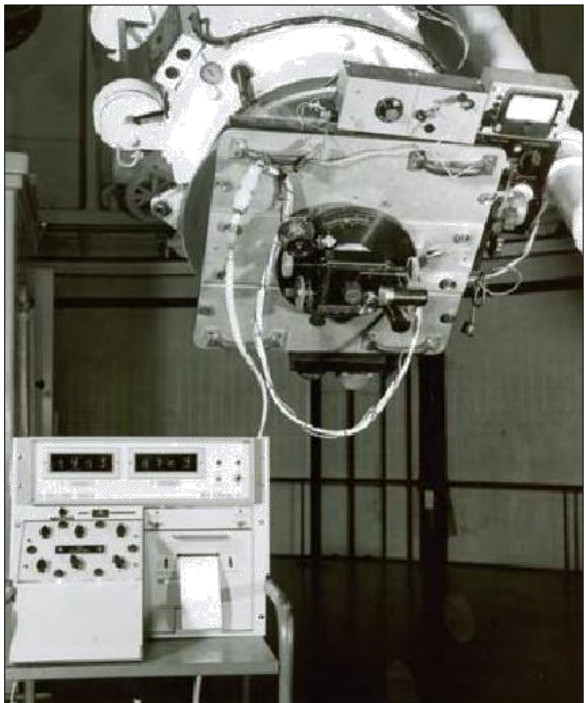

Figure 2: The micrometer, mounted on the 26-in refractor at the USNO that Charles Morley used for most of his observations (after Mason et al., 2007).

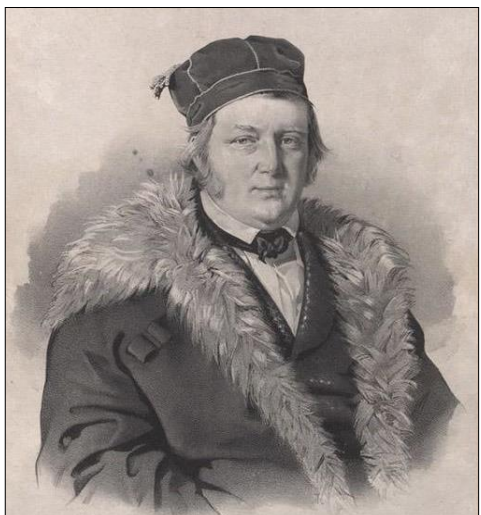

Figure 3: Wilhelm Struve (after Tenn, 2013: 83).

Based on a lifetime of experience, the late Charles Worley (1935–1997; Figures 1 and 2) acquired a very detailed knowledge of the observational uncertainties associated with individual observers, according to the time and number of the observations, the instruments they used, etc. Back in 1973 we were able to conduct extensive conversations with him and to profit from his knowledge of the reliability of the individual historical observations of separations (see Allen

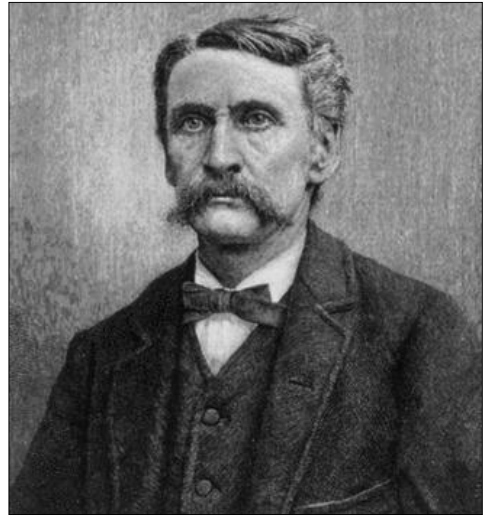

Figure 4: Sherburne Wesley Burnham (after Batten, 2014: 343).

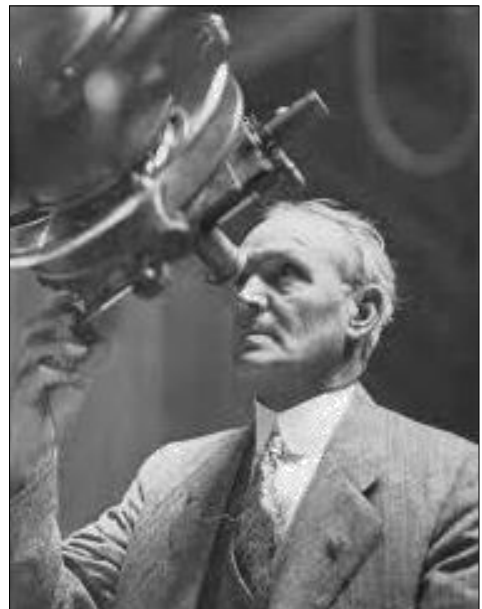

Figure 5: Robert Grant Aitken (after Jeffers, 1952: 4).

et al., 1974). He estimated that the uncertainties of the best observers were of about 0.07–0.08″ for Friedrich Georg Wilhelm von Struve (1793–1864; Figure 3), Sherburne Wesley Burnham (1838–1921; Figure 4) and Robert Grant Aitken (1864–1951; Figure 5); about 0.10″ for Willem Hendrik van den Bos (1896–1974), William Stephen Finsen (1905–1979) and George A. van Biesbrock (1880–1974); and about 0.12″ for Otto

Wilhem von Struve (1819–1905), Asaph Hall III (1829–1907), Ercole Dembowsky (1812–1881), George Cary Comstock (1855–1934), Paul Achille-Ariel Baize (1901–1995), William Joseph Hussey (1862–1926), Paul Couteau (1923–2014), Charles Edmund Worley and Joan George Arardus Gijsbertus Voûte (1879–1963). These uncertainties are sufficiently small to be meaningfully combined with more recent measures. Table 1 lists historical and modern observers along with their associated uncertainties (Allen et al., 2004).

Figure 6 shows the Orion Trapezium with the main components identified. Figure 7 shows the separation of some of the components of the Orion Trapezium as a function of time using historical and modern measures. For both historical and modern measures we used the data given in the Washington Double Star catalogue (WDS) (Mason et al., 2001) but we supplemented them with our results obtained with the diffracto-astrometry technique applied to Hubble Space Telescope images of the Orion Trapezium (for full details see Olivares et al., 2013).

Figures 8, 10 and 12 show the multiple systems ADS 719, ADS 13374 and ADS 15184 and their main components. Figures 9, 11, and 13 are examples of the separation velocities between some of the components of these multiple systems, obtained by combining historical with modern measures. These multiple systems resemble the Orion Trapezium, being composed of massive stars with similar separations. For the historical observations we used again the data listed in the WDS. Uncertainties for each modern observer cited in the WDS were estimated from differences in the separation measures conducted at nearly the same time. Separ-

Table 1: Errors of the most reliable observers

| Best (0.1″) | | Good (0.12″) |
|---|---|---|
| W. Struve | 0.07″ | O. Struve |
| Burnham | 0.07″ | Hall |
| Barnard | 0.08″ | Dembowsky |
| Aitken | 0.10″ | Comstock |
| van den Bos | 0.10″ | Baize (t > 1935) |
| Finsen | 0.10″ | Hussey |
| Van Biesbroek | 0.10″ | Couteau |
| USNO (t > 1960) | 0.04″ | Heintz |
| | | Worley |
| | | Voûte |
| All others: > 0.20″ | | |
| Modern Observers | | 0.01″ to 0.12″ |

ation measures among other pairs of components yielded similar figures. Note in each case the long time span covered by the observations.

## 3 THE DYNAMICAL EVOLUTION OF THE ORION TRAPEZIUM

To model the dynamical evolution of the trapezium-type systems we conducted Monte Carlo N-body simulations, that is, we numerically integrated the equations of motion of ensembles of fictitious systems representing each trapezium. As initial conditions for the Orion Trapezium we used the planar positions and the separation velocities as a lower bound for the transverse velocities. The distance to the Orion Trapezium, as well as the component masses and radial velocities were taken from the literature. Random perturbations were assigned to each quantity, with dispersions representative of the observational uncertainties. Since no information on the 'depth' of the system is available, random z-positions were assigned, within the projected radius of the system. Full details are given in Allen et al. (2017). Thus, we constructed a total of 300 fictitious clusters, which were integrated for one million years, with two values for the mass of Component C, namely 45 $M_\odot$ (the most recent value) and 65 $M_\odot$ (a plausible upper limit). Surprisingly, with the former value of the mass most systems dissolved in less than 10 thousand years. This is a very short time compared with the estimated evolutionary ages of the component stars, about a million years. However, the age of the Orion Nebula that surrounds the Trapezium is much shorter, of the order of 15 thousand years. Assuming the larger mass, we obtained longer lifetimes, about 40 thousand years. These longer lifetimes are compatible with the ones we previously obtained for the mini-cluster associated with Component B of the Orion Trapezium, also about 40 thousand years (Allen et al., 2015). Our simulations thus tend to favour the larger value for the mass of Component C.

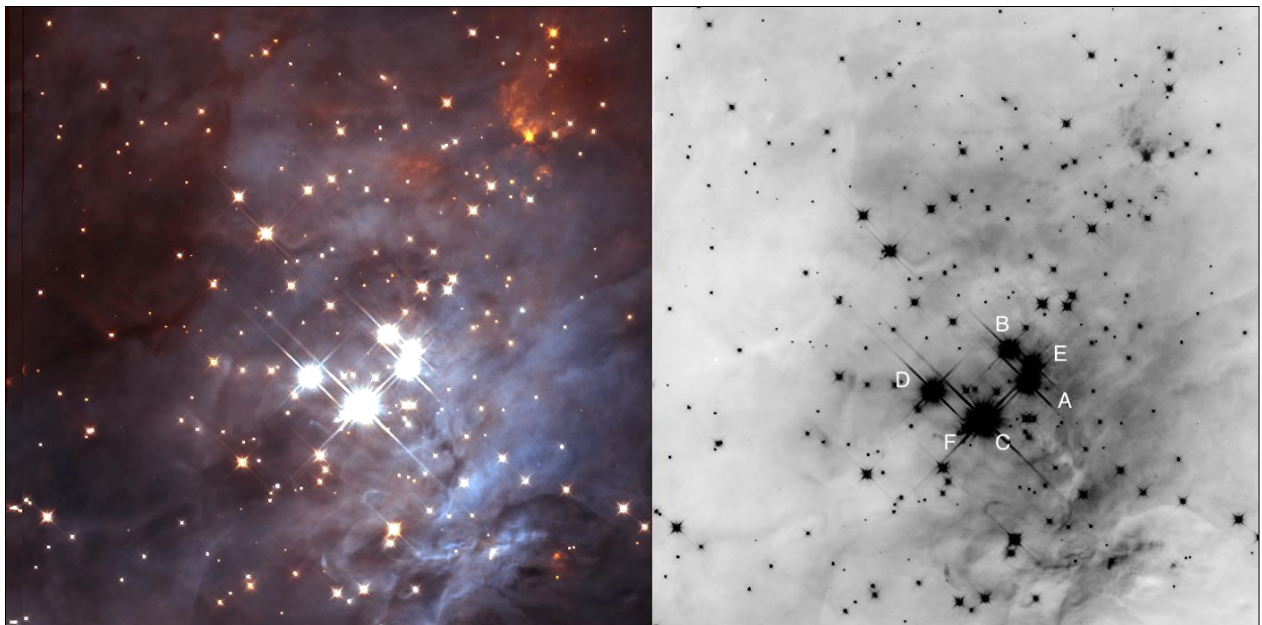

Figure 6: The Orion Trapezium Cluster. The six brighter components have been marked in the black and white negative image (right). Field of view is ~2.37 × 2.37 arcminutes. NASA-ESA/HST NICMOS Near-IR image.

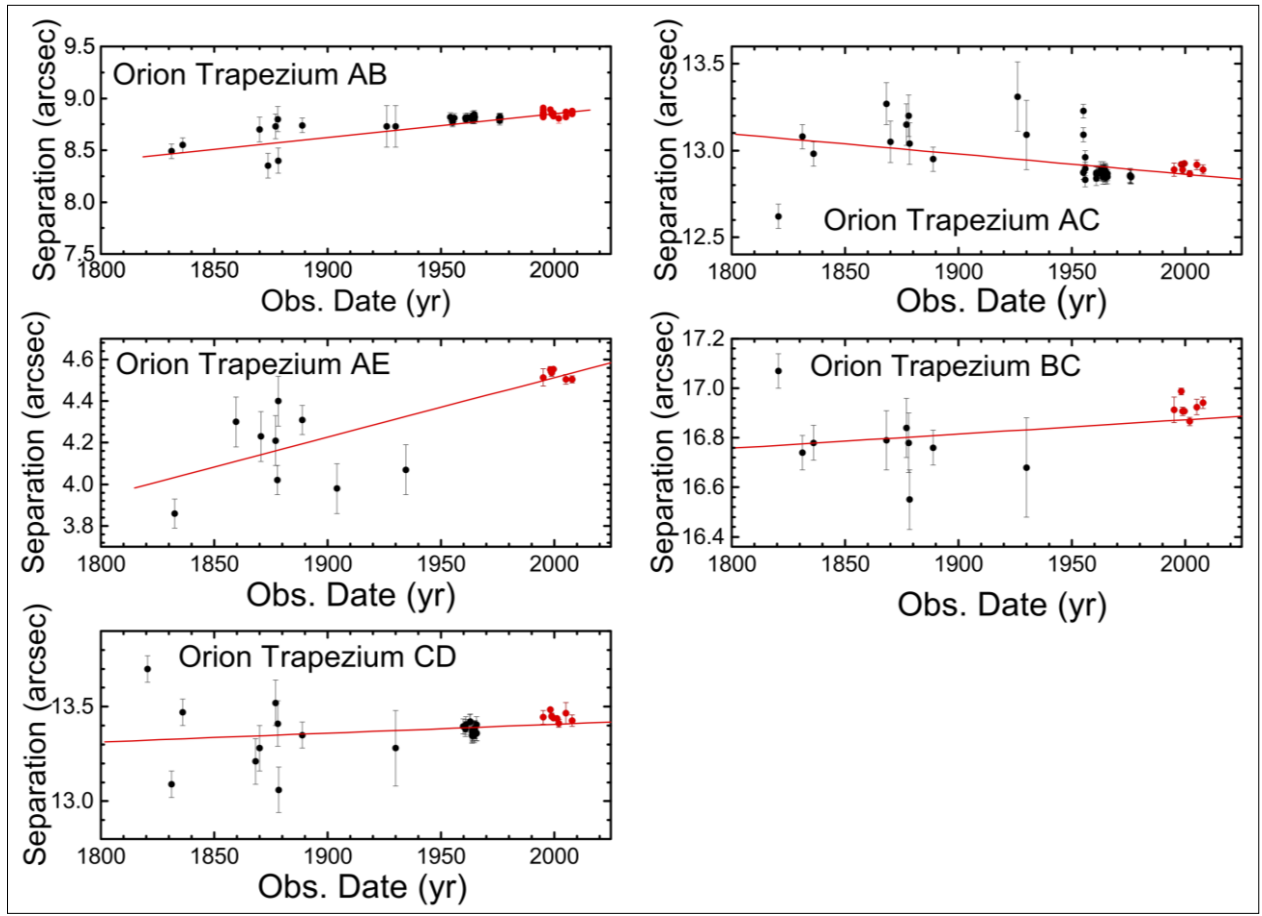

Figure 7: Separations of the Orion Trapezium components A relative to B, A to C, A to E, B to C and C to D as a function of time. Black circles correspond to WDS data with their associated error bars. Early measures by W. Struve, Burnham, Aitken, Barnard, van den Bos and Finsen are plotted. Red circles denote diffracto-astrometry HST data. Their associated error bars are smaller than the symbols. The lines show the best least-squares fit to the data.

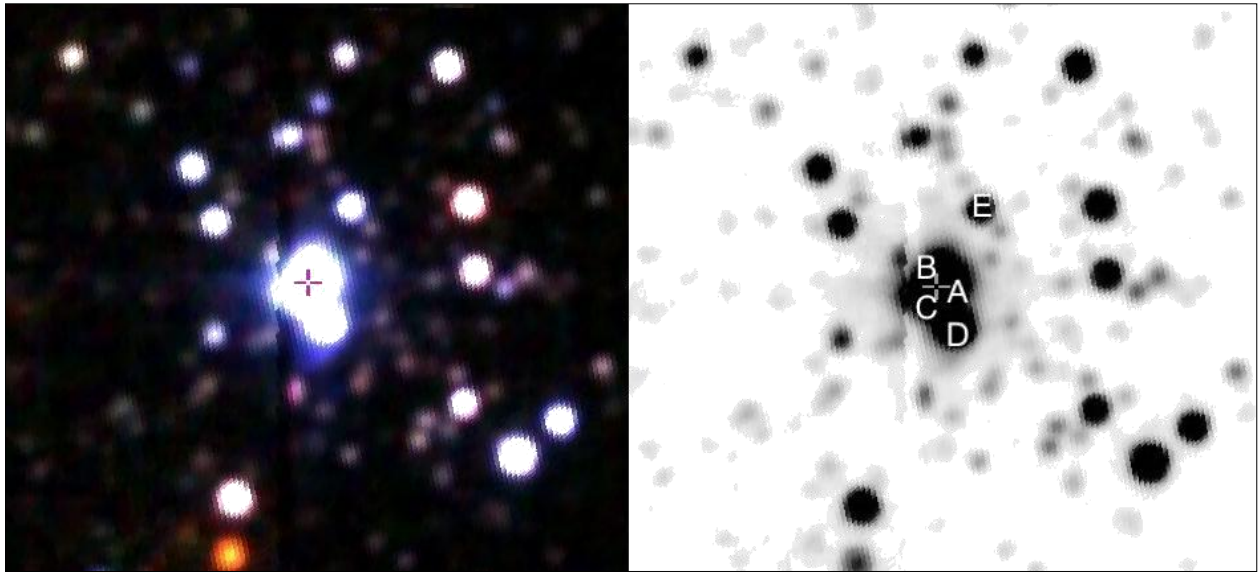

Figure 8: ADS 719. Identification of five components (A, B, C, D and E) have been marked in the black and white negative image (right). Field of view is ~2 × 2 arcminutes. North is up, East is to the left. 2MASS J, H, K bands composite image.

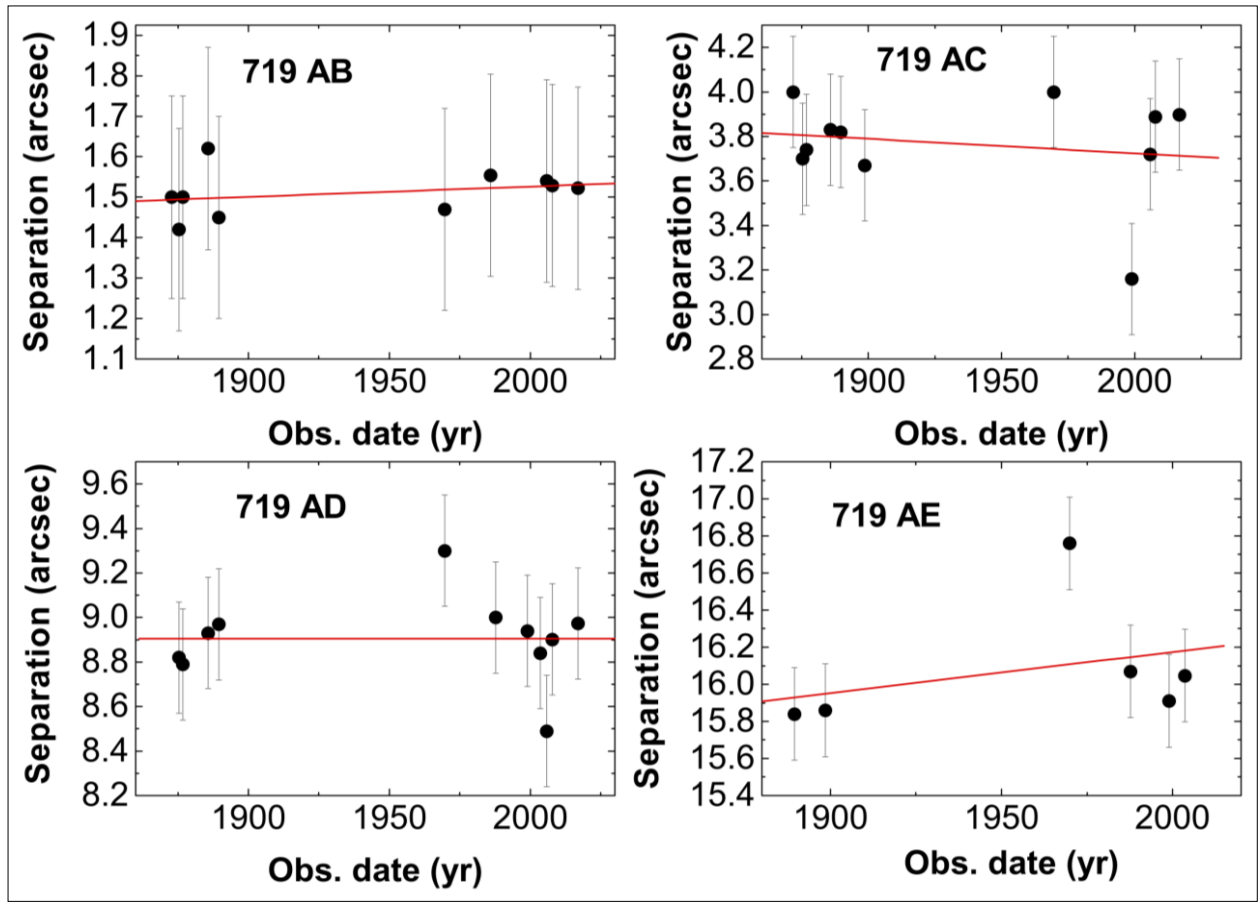

Figure 9: Example of separation measurements between pairs of components versus date of observation for the stars of trapezium ADS 719. Historical and modern data are combined. Early measures by Burnham, O. Struve, and Dembowsky are plotted. The line represents the least squares fit to the points from which we obtained the separation velocities.

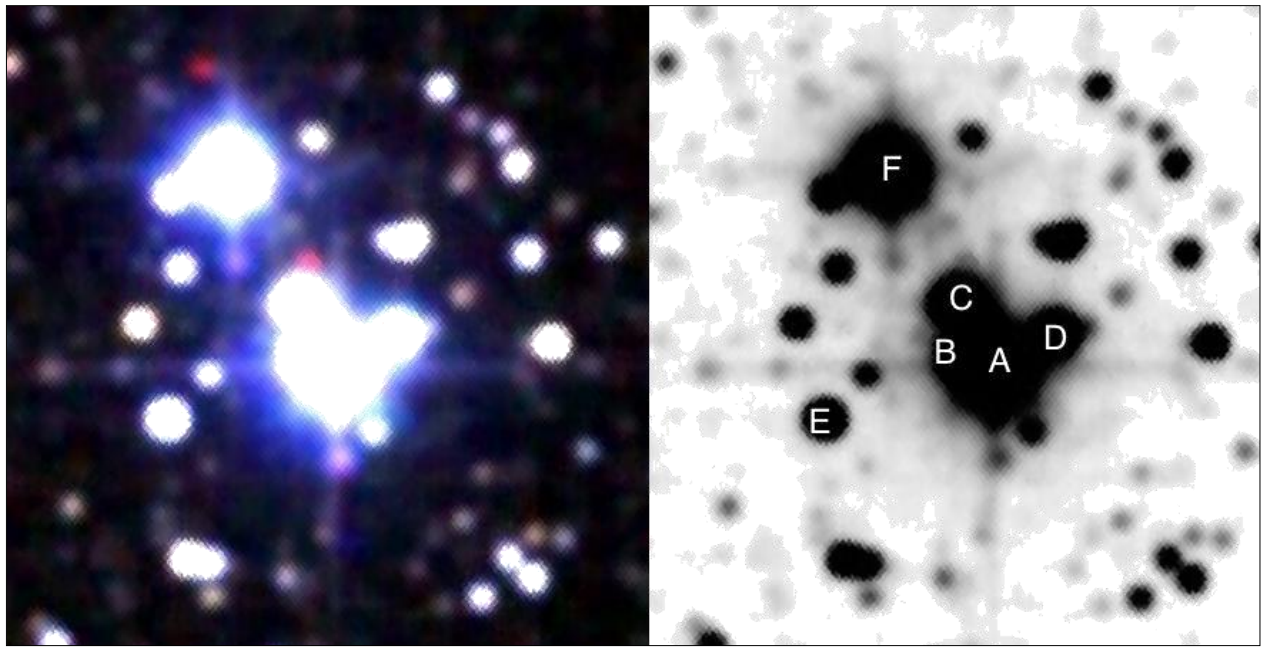

Figure 10: ADS 13374. Identification of six components (A, B, C, D, E and F) have been marked in the black and white negative image (right). Field of view is ~2.1 × 2.1 arcminutes. North is up, East is to the left. 2MASS J, H, K bands composite image.

Figure 14 shows the lifetimes and the types of systems resulting from the numerical simula-tions. As seen in the figure, the simulations produced a large number of binaries and triple stars, both hierarchical and non-hierarchical. Thus, the dynamical disintegration of systems resembling the Orion Trapezium will populate the field with massive binaries and triples. It is interesting to compare the properties of the binaries resulting from the fictitious clusters with those of observed field binaries. Most of the computed binaries have semi-axes of a few thousands AU, but binaries as close as a few AU were also formed. The frequency distributions of major semi-axes and eccentricities of the binaries formed during the numerical simulations do indeed resemble those of field binaries.

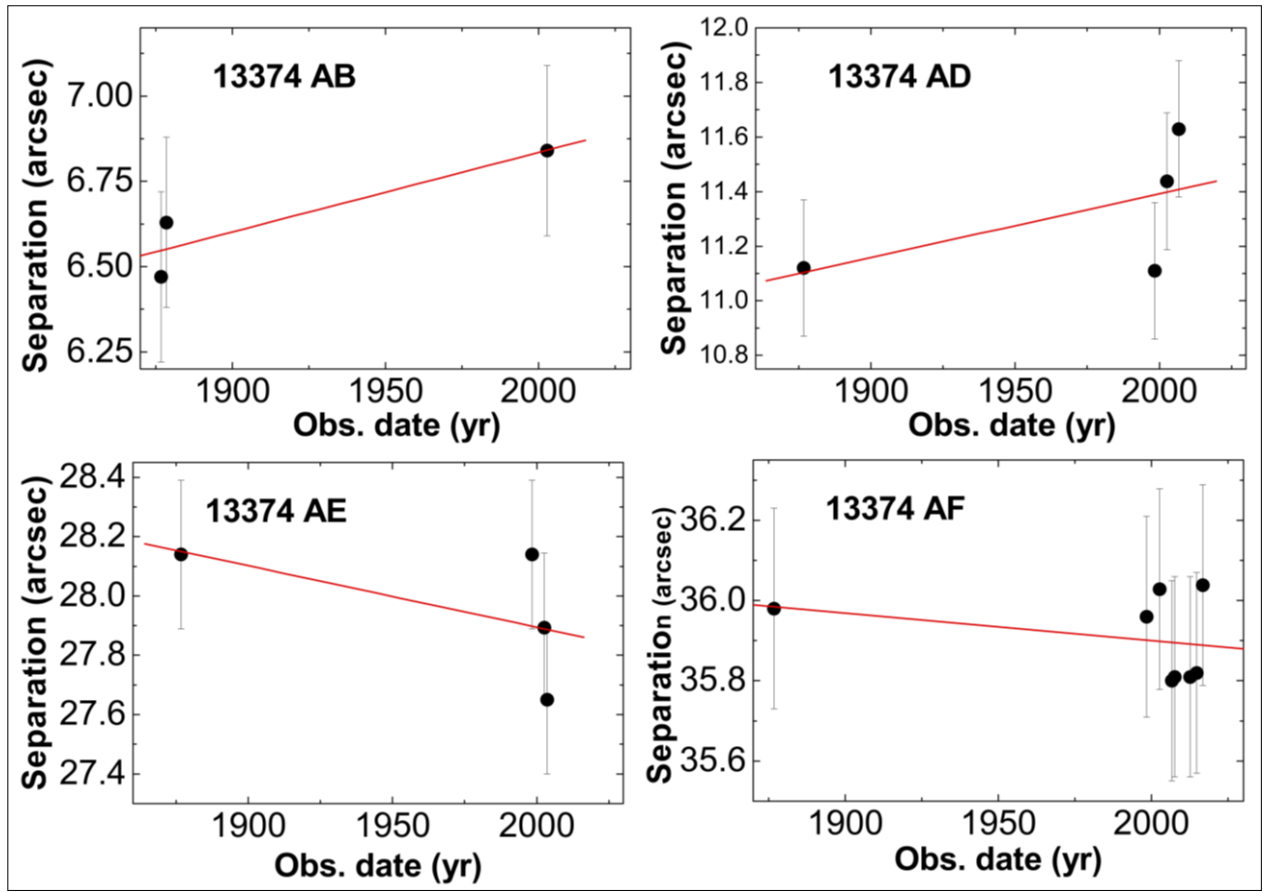

Figure 11: Same as for Figure 9 but for ADS 13374. Early measures by Burnham and Dembowsky are plotted.

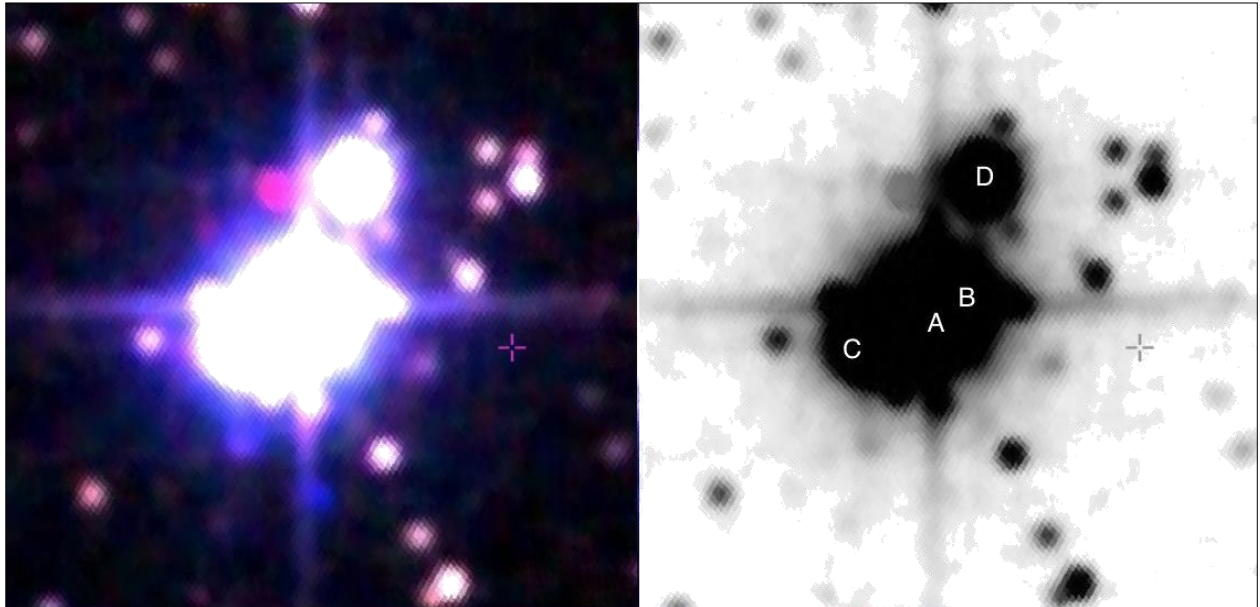

Figure 12: ADS 15184. Identification of four components (A, B, C and D) have been marked in the black and white negative image (right). Field of view is ~1.7 × 1.7 arc minutes. North is up, East is to the left. 2MASS J, H, K bands composite image.

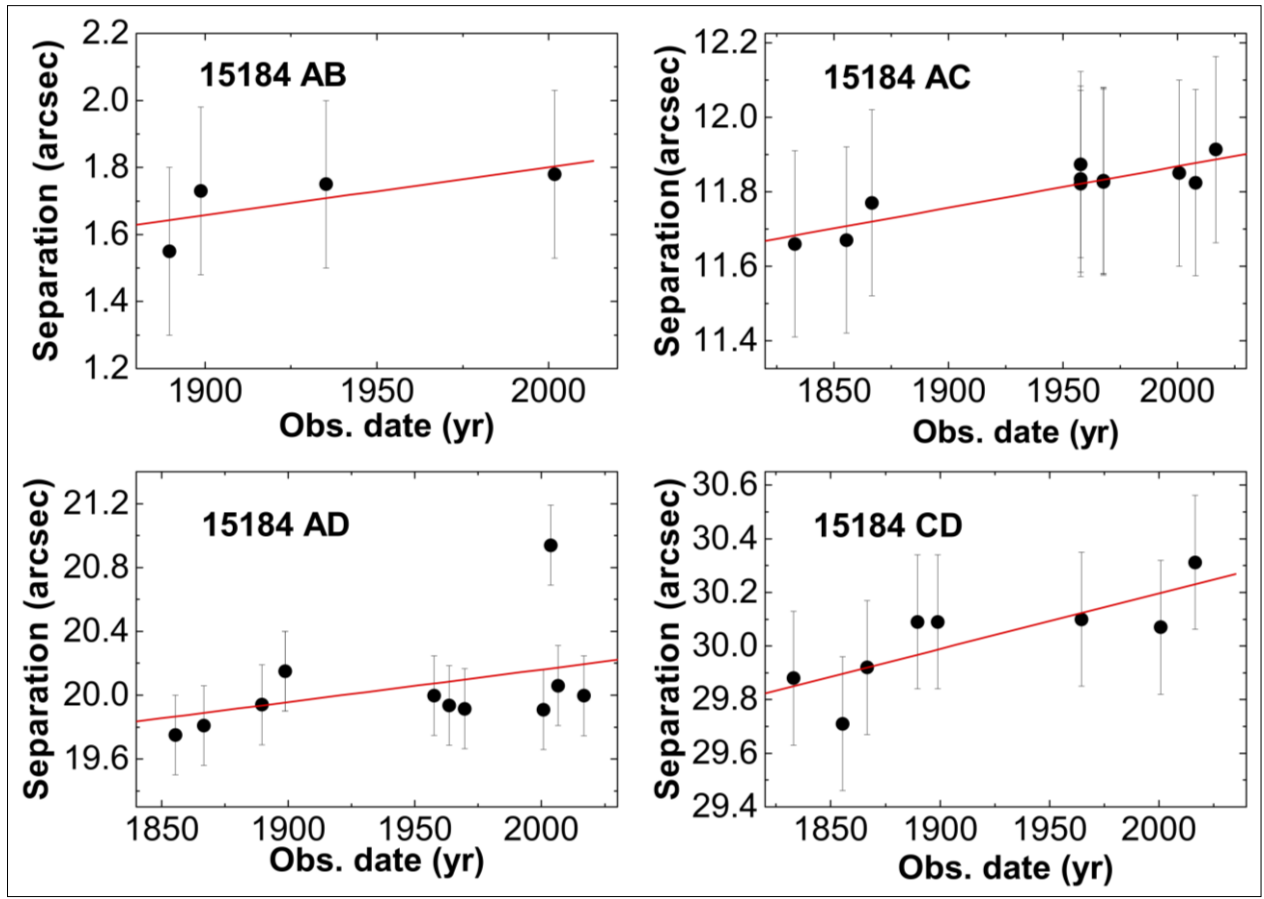

Figure 13: Same as for Figure 9 but for ADS 15184. Early measures by Dembowsky, Burnham and Aitken are plotted.

## 4   THE DYNAMICAL EVOLUTION OF TEN MASSIVE TRAPEZIA

We proceeded exactly as in the case of the Orion Trapezium. As initial conditions we took the planar positions and separation velocities, along with the best available data for the distances and masses. Again, no information on the 'depth' of these systems is available, so z-positions were randomly assigned. For the individual components of these systems we found no radial velocities in the literature, so we assigned them randomly, with values similar to the separation velocities. Perturbations representing the observational uncertainties were also applied.

Again, we generated a large number of fictitious clusters and performed N-body integrations of 100 perturbed systems representing each trapezium. For a more detailed explanation see Allen et al. (2018).

Figure 15 shows some results for ADS 719, one of the few resulting bound systems. Indeed, most of the simulated systems turned out to be unbound, even doubling the values of the component masses. We assumed twice the values for the masses to take into account the probable presence of undetected binaries among the components of the trapezia, since a large fraction of the massive stars are known to have companions. Doubling the masses should also generously allow for the possible gas fraction that escaped from the trapezium shortly after star formation. The bound systems were found to disintegrate in 10 to 20 thousand years.

In the majority of the simulated clusters the most massive star formed a binary, often accom-panied by the second most massive star. These binaries should end up as field binaries. The frequency distributions of major semi-axes and eccentricities of the binaries resulting from the disintegration of bound systems were found to be similar to those observed for field binaries. However, the binaries resulting from unbound systems do not accord as well with observations.

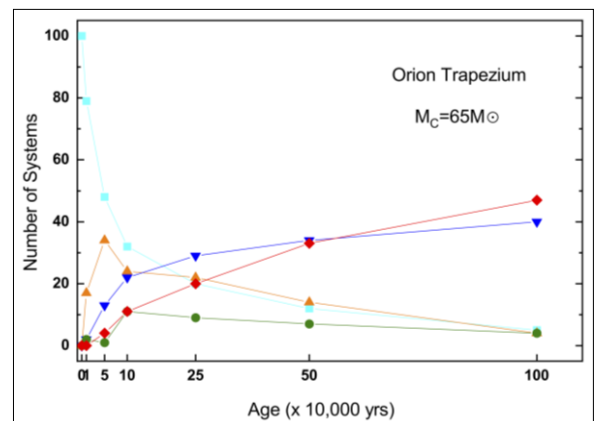

Figure 14: Numbers of different systems generated during the integrations of an ensemble of 100 fictitious clusters resembling the Orion Trapezium (mass of Component C: 65 $M_\odot$), as a function of age. Trapezia are shown as blue squares, hierarchical systems as green circles, non-hierarchical triples as orange triangles, hierarchical triples as blue inverted triangles, and binaries as red diamonds. The dynamical lifetime of the Orion Trapezium, defined as the time at which the number of systems decreases to one-half of the initial number, is about 40 thousand years.

## 5  CONCLUSIONS

We have shown that historical measures of the separations of the components of trapezium systems are important for studies of their dynamical evolution, since they can be combined with modern measures to provide information on the internal transverse velocities of the trapezium components. With knowledge of the transverse velocities provided by historical and modern mea-sures, and randomly assigning z-positions and radial velocities (when unavailable), dynamical evolution models can be constructed.

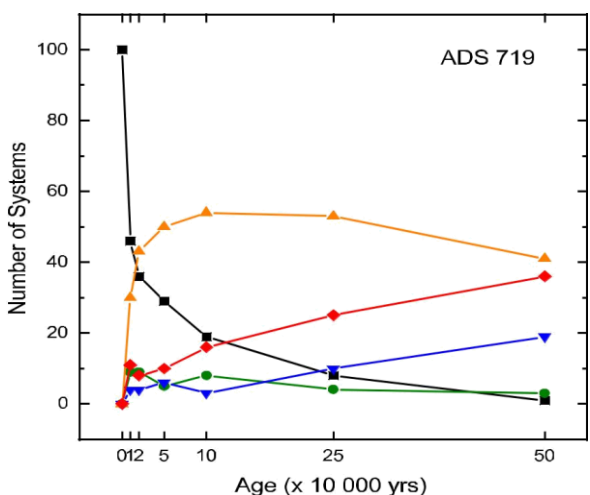

Figure 15: Numbers of different systems generated during the integrations of an ensemble of 100 trapezia resembling ADS 719, as a function of age. Trapezia of type 1, 2, 3, and 4 are shown as black squares, trapezia of type 1-2, 3, and 4 (with a tight binary) as green circles, non-hierarchical triples as orange triangles, hierarchical triples as blue inverted triangles, and binaries as red diamonds. The dynamical lifetime of ADS 719, defined as in Figure 14, is extremely short, about 10 to 20 thousand years.

The models obtained from a large number of simulated systems show that massive trapezia are dynamically very unstable. Indeed, most trapezia turn out to be gravitationally unbound; they disperse in less than 10 thousand years. Bound systems disintegrate in only 10 to 40 thousand years. These very short dynamical lifetimes are surprising if we compare them with the evolutionary lifetimes of the member stars. However, after a comprehensive literature search we found data for only 10 systems of spectral types O-B3, and thus likely to be true trapezia. Compared to the number of O-B3 stars from which the sample of 10 trapezia was obtained, this is a very small number. Thus, true trapezia are relatively scarce in the field, which is consistent with the very short dynamical lifetimes we found.

The end result of the numerical simulations was usually a wide binary, sometimes a triple of hierarchical or non-hierarchical type. Some hierarchical triples survive for up to the full integration time, a million years. Non-hierarchical triples survive for only up to 300 thousand years. The wide binaries formed will end up as field binaries. The binaries stemming from the bound trapezia have properties similar to those of observed wide binaries. Those formed in unbound systems do not accord well with observations.

With these examples as illustrations our hope has been to draw attention to the value of historical observations for modern research. There are veritable treasures of useful historical observations that are easily available, continue to be valuable, but remain largely ignored. Perhaps the examples we have provided will inspire further use of historical measures to obtain new insights.

## 6  ACKNOWLEDGEMENTS

We are grateful for the comments of three reviewers, which greatly helped to improve this paper. Our warmest thanks to E. Griffin, who strongly encouraged us to publish this work. L.J.S. and A.R-M thank DGAPA-UNAM for financial support under PAPIIT projects IN102517 and IN102617. This work has made use of the Washington Double Star Catalogue maintained at the U.S. Naval Observatory.

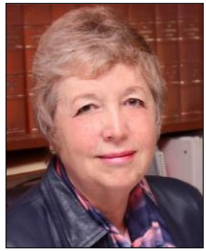

**Christine Allen** was born in Mexico. Since 1972 she has been a full time researcher at the Astronomy Institute of the National Autonomous University of Mexico, where she achieved tenure in 1978. Her main research interests are galactic structure, stellar dynamics, double and multiple stars, and star clusters. Christine developed in 1986 and 1991 a model for the mass distribution in the Milky Way, which has been widely used. She has compiled several catalogues of double and multiple systems. The latest, containing wide halo common proper-motion binaries formed the basis for a paper establishing that MACHOs (massive compact halo objects) must have masses smaller than 5 solar masses.

In recent years Christine has turned her attention to the dynamical evolution of young clusters. Still more recently, working with X. Hernandez, Christine has found evidence that in the very small acceleration regime, as found in the outer parts of globular clusters and in the widest binaries, Newtonian dynamics is not followed, a result that has been confirmed with Gaia data. The breakdown of classical gravity in the low acceleration regime has, of course, wide-ranging consequences.

Altogether, Christine has published over 100 research papers, which have gathered over 1000 citations. She has also published two books and over one hundred outreach articles in newspapers and magazines. Since 2001 she has served as Editor of *RMxAA* (*Revista Mexicana de Astronomía y Astrofísica*). Christine has lectured regularly. She has supervised 6 Bachelors theses, 2 Masters and 2 PhDs. Christine was a member of the now extinct IAU Commission 26 (Double and Multiple Stars). She was elected Vice-President in 2004 and became President in 2007.

Apart from her love of Astronomy, Christine loves classical music. She is active in several choirs. She particularly enjoyed singing Bach's Magnificat and B Minor Mass, Mozart's C Minor Mass and Vivaldi's Gloria.

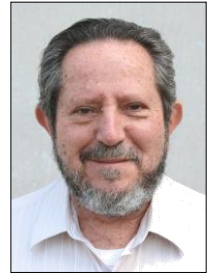

**Rafael Costero** was born in Mexico City in 1941, studied Physics at the Universidad Nacional Autónoma de México (UNAM) and obtained a Master degree in Astronomy from the University of Wisconsin, Madison. Until his recent retirement, he worked as a researcher at the Instituto de Astronomía, UNAM. On three occasions he was Assistant to the Director of this same Institute and was head of the Observatorio Astronómico Nacional at San Pedro Mártir (OAN-SPM), in Baja California. He has been visiting astronomer at the University of California, Santa Cruz, at the Osservatorio Astronomico di Trieste and at l'Observatoire de Marseille. His research work has covered areas such as planetary nebulae, interstellar medium, active galactic nuclei, hot stars and cataclysmic variables. Currently, he is investigating the bright stars in the Orion Trapezium. He also contributed to the instrumentation of the OAN-SPM. Formerly he taught basic Astronomy at the College of Geography at UNAM. Since a young boy, on and off, he has sung in different choirs, played the acoustic guitar and the bandurria, and serenaded several ladies.

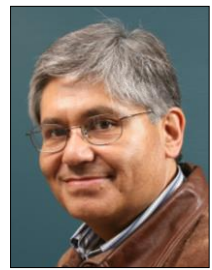

**Alex Ruelas-Mayorga** is a research scientist based at Instituto de Astronomía in the Universidad Nacional Autónoma de México (UNAM). He majored in physics at this university in 1977 and then took his Masters degree at the Victoria University of Manchester in the United Kingdom in 1979. He worked at the Nuffield Radio Astronomy Laboratories at Jodrell Bank. While at Jodrell Bank he worked under Professor Rod Davies and Dr Jim Cohen. In early 1979 he moved to Mt. Stromlo and Siding Spring Observatories in Canberra where he obtained his PhD in astronomy under Professor Harry Hyland and Dr Terry Jones, conducting research in galactic structure using the then new Mt. Stromlo Infrared Photometer. He has been a research scientist at Instituto de Astronomía since 1990 where he has conducted research in several fields of astronomy such as: stellar populations, globular clusters, elliptical and spiral galaxies, dynamics of stellar groups, cataclysmic variables and atmospheric turbulence. He has also been a lecturer at UNAM, and a theses supervisor in the undergraduate and graduate programmes of this university. (Picture taken by J. C. Yustis).

**Leonardo J. Sánchez Peniche** is a researcher at the Instituto de Astronomía in the Universidad Nacional Autónoma de México (UNAM) in Mexico City. He studied physics at the UNAM and subsequently completed his Masters and PhD at the University of Nice-

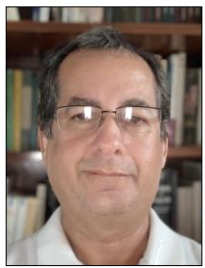
Sophia Antipolis, France. He works with collaborators in diverse areas of astrophysics such as: observational astronomy, astrometry, photometry, astronomical instrumentation and astronomical site evaluation. Leonardo has been part of the Atmospheric Site Evaluation group of the National Astronomical Observatory in San Pedro Mártir, whose results place this Observatory among the best astronomical sites in the world. He actively participates in the training of new astronomers, physicists and engineers through the teaching of courses and thesis direction both in the Faculty of Sciences and in the Astrophysics graduate programme at the UNAM.